# Do we have enough air?

V. Dorobantu*

Physics Department, Politehnica University, Timisoara, Romania

**Abstract**
*If only human beings would breathe the entire quantity of terrestrial air, then, at the present-day population on the Earth, one million years would exhaust that air. It seems we have enough air. Is it so?*

**Introduction**
This paper has a direct target of education. Recently, I have asked my students (polytechnic first year) to answer the question : do we have enough air? Essentially , how many air molecules the terrestrial atmosphere has. Towards my disappointment, I have received no appropriate answer. Let's see how do we deal with it.

**Method 1**
Let $p$ be the atmospheric pressure at the sea level. From the pressure definition, $p = \dfrac{F}{S}$, and knowing that the force F is due to the air weight, we will get the total mass of Earth's air :

$$m = \frac{p\,S}{g} \qquad (1)$$

The atmospheric pressure at sea level is 1 atmosphere = 101325 Pa, and the average gravitational acceleration $g = 9.80665$ m/s$^2$. The planet Earth is a prolate ellipsoid having the equatorial radius - $r_{ec} = 6378140$ m - as one axis, and the polar axis - $r_{pol} = 6356755$ m - , the other radius. The surface, and the volume, can be calculated using the formulae:

$$S = 2\pi\left(r_{ec}^2 + r_{pol}^2 \, \frac{\operatorname{arctanh}(e)}{e}\right) \text{ with } e = \sqrt{1 - \frac{r_{pol}^2}{r_{ec}^2}} \qquad (2)$$

$$V = 4\pi \frac{r_{ec}^2 \, r_{pol}}{3} \qquad (3)$$

---

E – mail: vdorobantu@gmail.com

If we consider the earth as a sphere with a radius of $r_{av} = 6367450$ m (the arithmetic mean of $r_{ec}$ and $r_{pol}$), we can calculate the surface using the expression:

$$S = 4\pi r_{av}^2 \qquad (4)$$

The volume will be
$$V = 4\pi \frac{r_{av}^3}{3} \qquad (5)$$

The air composition at $15\ ^0$C and 101325 Pa is:

| | |
|---|---|
| $N_2$ | 78.084 % |
| $O_2$ | 20.9476 % |
| Ar | 0.934 % |
| $CO_2$ | 0.0314 % |
| Ne | 0.001818 % |
| $CH_4$ | 0.0002 % |
| He | 0.000524 % |
| Kr | 0.000114 % |
| $H_2$ | 0.00005 % |
| Xe | 0.0000087 % |

Which means that we have to do with 28.9541 amu of air, or $m_0 = 4.80794 * 10^{-26}$ Kg. Using (1) and (4) we get the total mass of air: $5.26425 * 10^{18}$ Kg, and the total number of air molecule: $1.09491 * 10^{44}$. Using (1) and (2) we get $5.27014 * 10^{18}$ Kg, or $1.09613 * 10^{44}$ air molecules. As one can see, the results are very closed.

**Method 2**
Let us liquefy the air. As a result of liquefaction, on the entire surface of the Earth we will have a layer of liquid air of the height h. From $p = \rho g h$, with $\rho_{liquid\ air} = 880$ Kg/m$^3$ we obtain, h =11.7372 m. The volume of liquid air can be calculated using the formula:

$$V_{la} = \frac{4\pi}{3}\left[(r_{av} + h)^3 - r_{av}^3\right] \qquad (6)$$

or that one coming from (3). As we already have seen, the results are very similar. With $m = \rho_{liquid\ air} * V_{la}$ and $N = \frac{m}{m_0}$, we will get $1.09576 * 10^{44}$ molecules of air, almost the same number as the first method gave.

**Conclusions**
If the first method has given a huge number of air molecules, the second method gave the same result, but, in the same time, a very suggestive one: the entire quantity of air (liquid) is a layer of only 11.7372 m height. It is, generally, hard to imagine what means $10^{44}$, but easier, a height of almost 12 meters. So, do we have enough air? Not an easy answer!

Considering that, for an " average human being", six liters per minute of air circulate trough his lungs, then one million years are necessary to breath in the total number of air molecules. But, not only human beings breath. There are billions, or even trillions, of beings (or whatever) breathing in air. Fortunately, the air is recyclable. We ask ourselves if we breath in some number of molecules which already have been breathed by Einstein, for instance. Yes is the answer, situation which gives us some comfort hoping that those molecules of air will bring a seed of genius. But, what about the molecules breathed by Hitler, or Stalin, or who knows what monster? That one is not very comfortable.

*What about water?*

An estimation [1] of the total volume of water existing on Earth gives $1.386*10^{18}$ $m^3$, which means $1.386*10^{21}$ $kg$, or $4.63705*10^{46}$ $molecules$. So, the ratio of the number of water molecule and the number of air molecules is 463, roughly two order of magnitude. It does not seem to be very much. But, if we cover the Earth with entire quantity of water, we will get a layer of 2719 m height! Almost 3 Km! Here is a significance of two order of magnitude.